\documentclass{aa}
\usepackage{graphicx,natbib}
\bibpunct{(}{)}{;}{a}{}{,}

\newcommand{\nh}{N_\mathrm{H}}

\newcommand{\lesssim}{\mathrel{\hbox{\rlap{\hbox{\lower4pt\hbox{$\sim$}}}\hbox{$<$}}}}
\newcommand{\gtrsim}{\mathrel{\hbox{\rlap{\hbox{\lower4pt\hbox{$\sim$}}}\hbox{$>$}}}}
\newcommand{\beq}{\begin{equation}}
\newcommand{\eeq}{\end{equation}}
\newcommand{\beqa}{\begin{eqnarray}}
\newcommand{\eeqa}{\end{eqnarray}}

\begin{document}

\title{Identification of four RXTE Slew Survey sources with nearby
luminous active galactic nuclei}

\author{M.~Revnivtsev\inst{1,2} \and S.~Sazonov\inst{1,2} \and
E.~Churazov\inst{1,2} 
\and S.~Trudolyubov\inst{5,2}} 

\offprints{mikej@mpa-garching.mpg.de}

\institute{Max-Planck-Institut f\"ur Astrophysik,
           Karl-Schwarzschild-Str. 1, D-85740 Garching bei M\"unchen,
           Germany
     \and   
           Space Research Institute, Russian Academy of Sciences,
           Profsoyuznaya 84/32, 117997 Moscow, Russia
     \and       
           Institute for Geophysics and Planetary Physics, 
           University of California, Riverside, CA 92521, USA
}
\date{Received / Accepted}

\authorrunning{Revnivtsev et al.}
\titlerunning{Identification of four XSS sources}

\abstract{Based on RXTE scans and observations with the
SWIFT/XRT telescope and INTEGRAL observatory, we report the
identification of four X-ray sources discovered during the RXTE Slew
Survey of the $|b|>10^\circ$ sky with nearby ($z\sim$0.017--0.098)
luminous ($L_{\rm 2-10~keV}\sim 5\times
10^{42}$--$10^{44}$~erg~s$^{-1}$) active galactic nuclei. Two of the
objects exhibit heavily intrinsically absorbed 
X-ray spectra ($\nh\sim 10^{23}$~cm$^{-2}$). 
}
\maketitle

\section{Introduction}
\label{intro}

Shallow large-area X-ray surveys are essential for studying the local 
population of active galactic nuclei (AGN) and complement deep
small-area X-ray surveys, which probe AGN at high redshift. Recently, a
significant progress has been made in scanning the extragalactic sky
at photon energies above 2~keV, where circumnuclear obscuration 
does not affect the detectability of nearby AGN as severely as in the
soft X-ray regime. In particular, a serendipitous survey of
the high Galactic latitude ($|b|>10^\circ$) sky in the 3--20~keV band
was performed with the PCA instrument of the RXTE observatory -- the
RXTE Slew Survey (XSS, \citealt{rsj+04}), in which a total of $\sim
300$ sources with fluxes down to $\sim$0.5--1~mCrab were detected,
including $\sim 100$ identified AGN. This survey is now being nicely
complemented by hard X-ray (above 15~keV) all-sky surveys
performed with INTEGRAL (e.g. \citealt{bgs+05,scr+05}) and SWIFT
\citep{mts+05}, which are somewhat more sensitive to strongly obscured 
($\nh\gtrsim 10^{23}$~cm$^{-2}$) AGN compared to the XSS. 

About 30 XSS sources remain unidentified (most of them
located in the southern hemisphere), and this presents some
uncertainty for statistical studies based on the XSS AGN sample
\citep{sr04,hph+05}. The main reason for the missing identifications
is the poor localization of XSS sources. The positions of weak
XSS sources are known only to $\sim 60^\prime$ accuracy,
which practically precludes their identification through observations
with X-ray and optical telescopes. 

A number of steps have therefore been undertaken to improve the completeness of
identification of the XSS source catalog. First, several unidentified XSS
sources were recently detected in hard X-rays with INTEGRAL, resulting
in sufficiently accurate positions ($\sim 5^\prime$) for initiating
observations of these sources with Chandra. This led to the identification
of two sources (XSS J12389$-$1614 and XSS J19459+4508) with nearby
galaxies, implying that they are AGN \citep{scr+05}. Secondly, we
obtained RXTE/PCA scans over the positions of a set of XSS sources,
which allowed us to improve these positions to 6--20$^\prime$. Some of
these sources together with a few other XSS sources for which there is
a likely soft X-ray counterpart from the ROSAT all-sky survey (RASS,
\citealt{vab+99}) were recently observed by the X-Ray Telescope (XRT)
aboard SWIFT, providing accurate localizations and X-ray spectral
information. Finally, XSS sources with relatively well known positions
as well as INTEGRAL detected sources make up the input sample for an
ongoing program of optical identification of northern objects on the
Russian-Turkish 1.5-m Telescope \citep{bsr+06}.

The continuation of all these efforts should soon lead to a highly
complete identification of the XSS catalog. Here we report the
association of 4 previously unidentified XSS sources with nearby
luminous AGN, based on RXTE, SWIFT and INTEGRAL observations.

\section{Observations}

\begin{table*}
\caption{Identification of XSS sources 
\label{ident_table}
}

\begin{tabular}{lrrccll}
\hline
\hline
\multicolumn{3}{c}{RXTE} &
\multicolumn{1}{c}{Identification} &
\multicolumn{1}{c}{AGN} &
\multicolumn{1}{c}{Position$^{\rm b}$} &
\multicolumn{1}{c}{Redshift}
\\
\cline{1-3}
\multicolumn{1}{c}{Name} &
\multicolumn{1}{c}{$\alpha$, $\delta$ (2000)} &
\multicolumn{1}{c}{Err} &
\multicolumn{1}{c}{} &
\multicolumn{1}{c}{Class$^{\rm a}$} &
\multicolumn{1}{c}{$\alpha$, $\delta$ (2000)} &
\multicolumn{1}{c}{}
\\
\hline
XSS J05054$-$2348 & 76.414 $-$23.906 &  6$^\prime$ & 
 2MASX J05054575$-$2351139 & Sy2 & 05 05 45.7 $-$23 51 14 & 0.0350$^{\rm c}$\\
XSS J12303$-$4232 & 187.567 $-$42.541 & 60$^\prime$ & 
 1RXS J123212.3$-$421745 & & \\
 & & &  IRAS F12295$-$4201? &  &  \\
 & & &  USNO-B1.0 0477-0336688 & & 12 32 12.0 $-$42 17 51 & 0.098$^{\rm d}$ \\
XSS J16151$-$0943 & 243.796 $-$09.724 & 60$^\prime$ & 
 1RXS J161519.2$-$093618 & & \\
 & & & 6dF J1615191$-$093613 & Sy1 & 16 15 19.1 $-$09 36 13 & 0.0650$^{\rm c}$ 
\\
XSS J18236$-$5616 & 275.703 $-$56.348 & 20$^\prime$ & 
 IGR J18244$-$5622 & &\\
 & & &  IC 4709 & & 18 24 19.4 $-$56 22 09 & 0.0169\\
\hline
\end{tabular}

$^{\rm a}$ From \cite{bsr+06}

$^{\rm b}$ The quoted optical positions are consistent with those
measured by SWIFT/XRT or INTEGRAL/IBIS

$^{\rm c}$ The quoted NED values are confirmed by \cite{bsr+06} 

$^{\rm d}$ Inferred from the X-ray spectrum 

\end{table*}

Dedicated RXTE/PCA scans performed on June 23, 2004 and March 28, 2005
allowed us to significantly improve the positions of two sources,
XSS J05054$-$2348 and XSS J18236$-$5616. In two other cases, XSS
J12303$-$4232 and XSS J16151$-$0943, a likely bright ($\sim 0.6$ and $\sim 0.4$
cnt~s$^{-1}$, respectively) soft X-ray counterpart from the RASS was
previously proposed based on the original XSS localization
\citep{sr04}, with the positions of the RASS sources known to $\sim
10''$ accuracy. During the period August--October 2005 three of the
aformentioned sources (XSS J05054$-$2348, XSS J12303$-$4232 and XSS
J16151$-$0943) were observed with the SWIFT/XRT telescope for exposure
times ranging from 3 to 26~ks.

\begin{figure*}
\centering
\includegraphics[bb=0 0 590 530, width=0.70\textwidth]{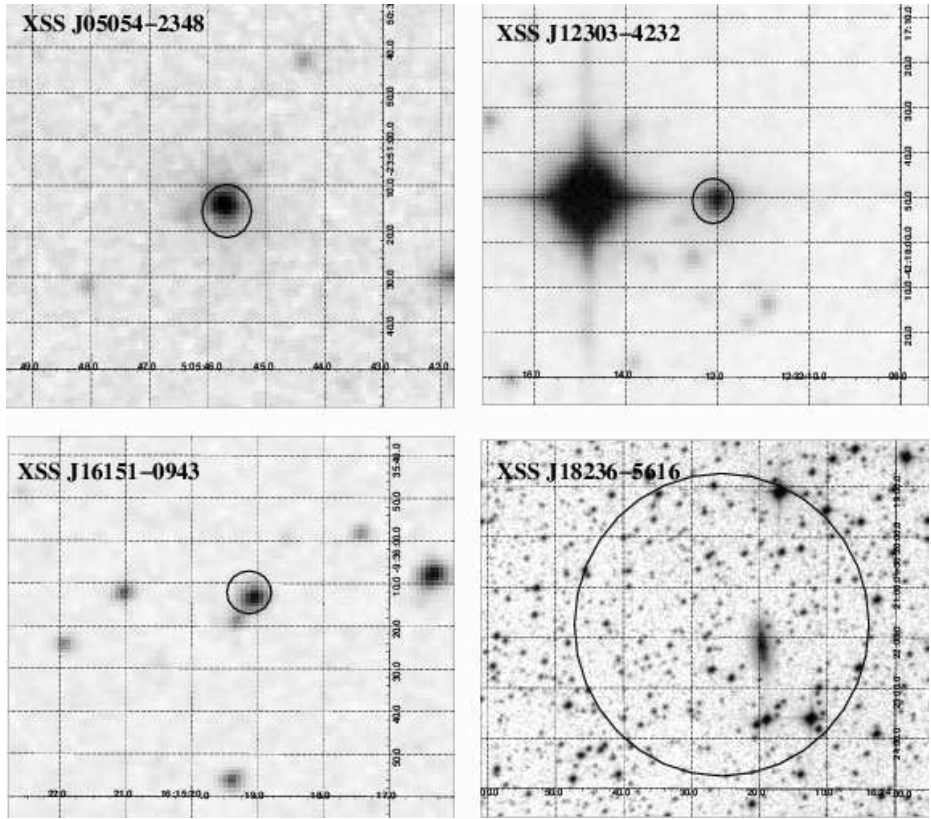}
\caption{SWIFT/XRT localizations ($\sim 5''$-radius circles) and
an INTEGRAL/IBIS 1$\sigma$ confidence region ($\sim 
3^\prime$-radius circle in the lower right panel) for the studied XSS sources,
plotted against the DSS optical images.
}
\label{images_fig}
\end{figure*}

We analyzed the SWIFT/XRT observations using standard tasks of
HEADAS 6.0.1 package following SWIFT Guest Observer Facility recommendations
(http://legacy.gsfc.nasa.gov/docs/swift/analysis/) and found a single  
bright X-ray (0.5--10~keV) source in the $\sim 20^\prime\times 20^\prime$
XRT image for each XSS source. For XSS J12303$-$4232 and XSS
J16151$-$0943 the XRT detected source coincides with the proposed RASS
counterpart. If more than one X-ray source was present in the XRT
field of view, the positions of weaker X-ray sources were correlated
with the optical and infrared images of the region to calculate
astrometric corrections to the XRT image. By applying this
correction we were then able to improve the positions of the bright
X-ray sources to $\sim$1--2'' (without astrometric correction the
localization accuracy is $\sim 5''$). Cross-correlation of the XRT
images with optical sky surveys (see Table~\ref{ident_table} and 
Fig.~\ref{images_fig}) revealed that XSS J05054$-$2348 and XSS
J16151$-$0943 are each associated with a known galaxy. Optical
spectroscopy further demonstrated \citep{bsr+06} that these galaxies can
be classified as Seyfert 2 and Seyfert 1, respectively. 

The XRT position of XSS J12303$-$4232 coincides with a relatively
bright ($R\sim 14$) stellar-like object (see Fig.~\ref{images_fig}),
which has a likely infrared counterpart from the IRAS catalog (see
Table~\ref{ident_table}). As will be shown below, the X-ray spectrum
of XSS J12303$-$4232 obtained with XRT is typical for AGN and
contains a strong emission line at $\sim 5.8$~keV that can be
identified with a redshifted $K\alpha$ line of neutral iron. These
facts suggest that XSS J12303$-$4232 is a nearby QSO at $z\sim 0.1$. 

The fourth source of the present sample (XSS J18236$-$5616) has not
yet been observed by SWIFT/XRT but is clearly detectable (at
3.5$\sigma$) on the summed hard X-ray map accumulated by INTEGRAL/IBIS
so far (Revnivtsev et al., in preparation). This yields us a
significantly improved position ($\sim 5^\prime$) for this source. The
INTEGRAL confidence region contains a known nearby galaxy 
(see Table~\ref{ident_table} and Fig.~\ref{images_fig}), which
together with the broad-band X-ray spectrum taken by RXTE and INTEGRAL
(see below) strongly suggests that this source too is an AGN.

\subsection{X-ray spectra}

\begin{figure*}
\centering
\includegraphics[bb=0 160 580 720, width=0.72\textwidth]{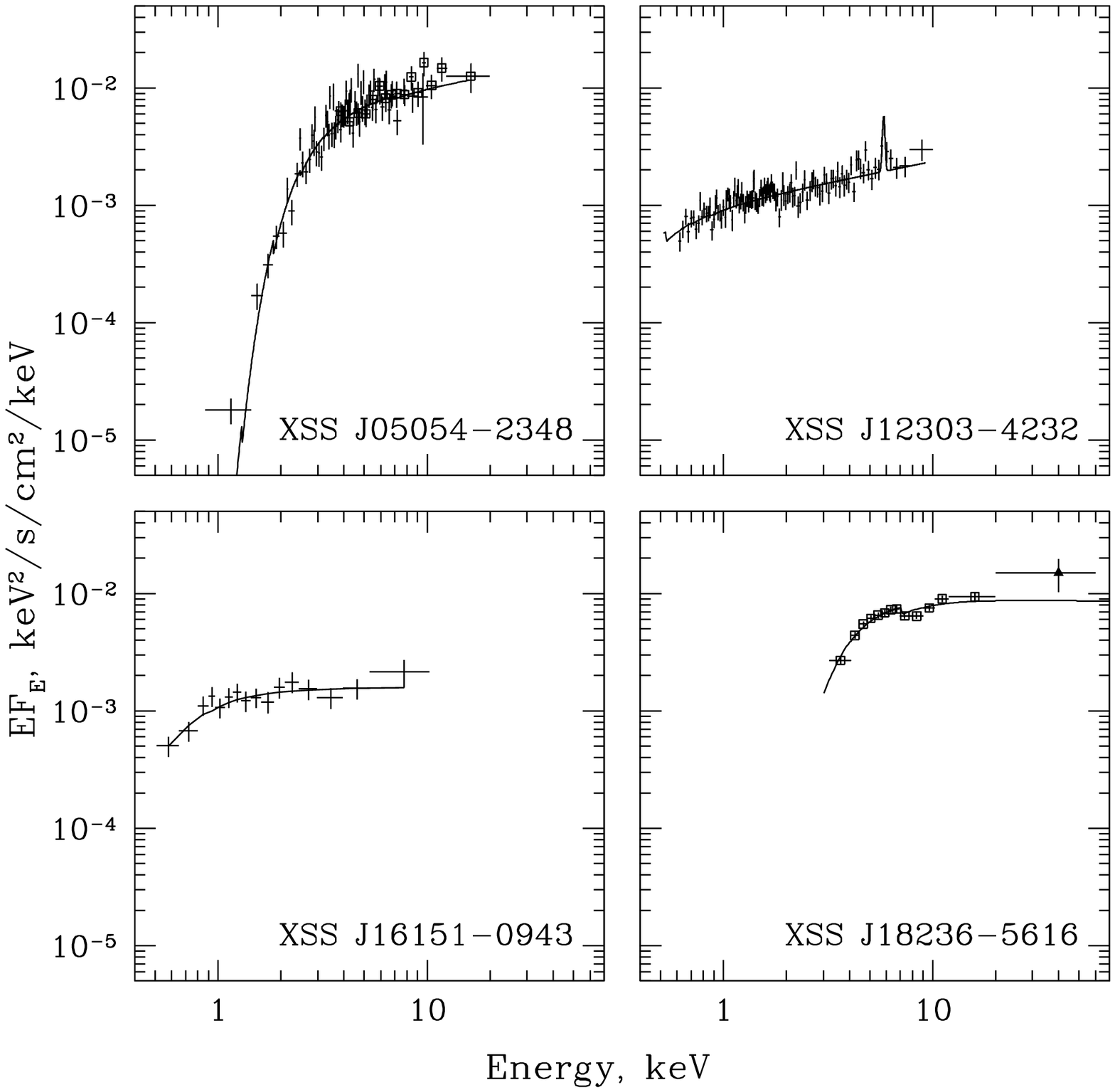}
\caption{X-ray spectra of the XSS sources. Shown are data from
SWIFT/XRT (between 0.5 and 10~keV for XSS J05054$-$2348, XSS J12303$-$4232
and XSS J16151$-$0943) and from RXTE/PCA (between 3 and 20~keV for XSS
J05054$-$2348 and XSS J18236$-$5616, open squares). For XSS
J18236$-$5616 also the 17--60~keV flux measured by INTEGRAL is
shown (solid triangle). The solid lines represent the best-fit models
described in Table~\ref{spec_table}.
}
\label{spectra_fig}
\end{figure*}

We then performed a 
spectral analysis based on SWIFT/XRT data in the 0.5--10~keV band for
XSS J05054$-$2348, XSS J12303$-$4232 and XSS J16151$-$0943, and
RXTE/PCA scans data in the 3--20~keV band for XSS J05054$-$2348 and
XSS J18236$-$5616. For the last source, there is also a
measurement of its 17--60~keV flux by INTEGRAL. Combining all these data
we obtain the source spectra presented in Fig.~\ref{spectra_fig}. 

\begin{table*}
\begin{center}
\caption{Results of spectral fitting, source fluxes and luminosities
\label{spec_table}
}

\begin{tabular}{ccccccccc}
\hline
\hline
\multicolumn{1}{c}{Source} &
\multicolumn{1}{c}{Instrument} &
\multicolumn{1}{c}{$\Gamma$} &
\multicolumn{1}{c}{$\nh$ ($N_\mathrm{H, Gal}$$^{\rm a}$)} &
\multicolumn{1}{c}{Line} &
\multicolumn{1}{c}{$\chi^2$/} &
\multicolumn{1}{c}{$F$ (2--10 keV)} &
\multicolumn{1}{c}{$L$ (2--10 keV)$^{\rm b}$}
\\
\multicolumn{1}{c}{} &
\multicolumn{1}{c}{} &
\multicolumn{1}{c}{} &
\multicolumn{1}{c}{$10^{22}$~cm$^{-2}$} &
\multicolumn{1}{c}{$E$ (keV)} &
\multicolumn{1}{c}{d.o.f} &
\multicolumn{1}{c}{$10^{-11}$~erg/cm$^{2}$/s} &
\multicolumn{1}{c}{$10^{43}$~erg/s}
\\
\hline
XSS J05054$-$2348 & XRT+PCA & $1.71\pm 0.12$ & $5.1\pm 0.4$ (0.02)
  & -- & 0.98 & $1.5\pm 0.1$ & $3.7\pm 0.2$ \\  
XSS J12303$-$4232 & XRT     & $1.63\pm 0.07$ & $0.05\pm 0.02$ (0.07)
  & $5.82\pm 0.03$ & 0.95 & $0.46\pm 0.04$ & $9.8\pm 0.9$ \\  
XSS J16151$-$0943 & XRT     & $1.98\pm 0.15$ & $0.14\pm 0.05$ (0.13)
  & -- & 0.67 & $0.40\pm 0.04$ & $3.6\pm 0.4$ \\  
XSS J18236$-$5616 & PCA     & $2.00\pm 0.13$ & $12\pm 2$ (0.08)
 & -- & 0.73 & $1.1\pm 0.1$ & $0.62\pm 0.06$ \\  
\hline
\end{tabular}
\end{center}
$^{\rm a}$ Galactic absorption column based on \cite{dl90} 

$^{\rm b}$ Observed luminosity, assuming $(H_0, \Omega_{\rm m},
\Omega_\Lambda)= (75$~km~s$^{-1}$~Mpc$^{-1}$, 0.3, 0.7)

\end{table*}
 
All four spectra are well fit by a power law modified by
photoabsorption along the line of sight (see
Table~\ref{spec_table}). The photon indexes are well constrained by
the data and are typical for Seyfert galaxies and QSOs ($\Gamma\sim
1.8$). The spectra of XSS J05054$-$2348 and XSS J18236$-$5616 exhibit 
heavy photoabsorption ($\nh\sim 10^{23}$~cm$^{-2}$), while in the
other two cases (XSS J12303$-$4232 and XSS J16151$-$0943) the measured
absorption column density is consistent with the Galactic absorption
toward the sources. The high and low absorpiton columns measured for
XSS J05054$-$2348 and XSS J16151$-$0943 are in accord with their
optical classification as Seyfert 2 and Seyfert 1 galaxy,
respectively. We also note the perfect agreement between the spectra
of XSS J05054$-$2348 obtained by SWIFT/XRT and RXTE/PCA. Similarly 
the hard X-ray flux measured with INTEGRAL for XSS J18236$-$5616 is
consistent (within the 90\% uncertainties) with the extrapolated
best-fit model of RXTE/PCA data.  

Very interesting is the case of XSS J12303$-$4232, whose spectrum
contains a bright narrow emission line at energy $\sim 5.8$~keV
(detected at $>99.9\%$ confidence). It is natural to interpete this
feature as a redshifted 6.4~keV iron fluorescence line, often observed
in high-quality AGN spectra. The measured equivalent line width
($280\pm40$~eV) is also typical for relatively unobscured AGN
\citep[e.g.][]{nandra94,turner97}. The determined line position then
suggests that XSS J12303$-$4232 is an AGN at a redshift of $\sim 0.1$. 

The determined luminosities of our objects (see
Table~\ref{spec_table}) are typical for Seyfert galaxies and fall near
the peak of the X-ray luminosity function of local AGN \citep{sr04}.  

\section{Conclusion}

We presented evidence that an additional 4 XSS sources are nearby
($z=0.017$--0.098) AGN. Two of them are characterized by strong
intrinsic absorption ($\nh\sim 10^{23}$~cm$^{-2}$). It is interesting
to note that this finding continues the tendency of finding
obscured objects almost exclusively on the low-luminosity ($\lesssim
10^{43.5}$~erg~s$^{-1}$) branch of the X-ray luminosity function of
local AGN, as first clearly noticed in the XSS \citep{sr04} and now
also in the SWIFT and INTEGRAL surveys \citep{mts+05,scr+05}.  

This study fills some of the remaining gaps in the XSS catalog. We
hope that the final catalog, consisting of up to $\sim 130$ X-ray
bright ($\gtrsim$0.5--1~mCrab), mostly nearby ($z<0.1$), AGN, will
serve as a useful input sample for detailed studies of the local AGN
population. In addition, efforts put into the identification of XSS
sources also help the continuing INTEGRAL and SWIFT hard X-ray all-sky surveys,
since there is expected to be significant overlap between all three
final catalogs of detected sources.

\smallskip
\noindent {\sl Acknowledgments} This research has made use of data
obtained through the High Energy Astrophysics Science Archive Research
Center Online Service, provided by the NASA/Goddard Space Flight
Center. This research has also made use of the SIMBAD database (operated at
CDS, Strasbourg) and NASA/IPAC Extragalactic Database (operated by the
Jet Propulsion Laboratory, California Institute of Technology).


\end{document}